\documentclass[twoside]{ilcws10}
\usepackage[latin1]{inputenc}
\usepackage[dvips]{graphicx,epsfig,color}
\usepackage{wrapfig,rotating}
\usepackage{amssymb,amsmath,array}

\newcommand{\zh}{$Z_{\mathrm{H}}$}
\newcommand{\ah}{$A_{\mathrm{H}}$}
\newcommand{\wh}{$W_{\mathrm{H}}$}

\newcommand{\zhah}{$A_{\mathrm{H}}Z_{\mathrm{H}}$}
\newcommand{\whwh}{$W_{\mathrm{H}}^{+}W_{\mathrm{H}}^{-}$}
\newcommand{\zhzh}{$Z_{\mathrm{H}}Z_{\mathrm{H}}$}
\newcommand{\zhzhahahbbbb}{$Z_{\mathrm{H}}Z_{\mathrm{H}} \to A_{\mathrm{H}}A_{\mathrm{H}} bbbb$}

\begin{document}
\title{Analysis of Little Higgs Model with T-parity at ILC} 
\author{Yosuke Takubo$^1$, 
Eri Asakawa$^2$,
Masaki Asano$^1$,
Keisuke Fujii$^3$,
Eriko Kato$^1$, \\
Shigeki Matsumoto$^4$, and
Hitoshi Yamamoto$^1$
\vspace{.3cm}\\
1- Department of Physics, Tohoku University, Sendai, Japan \\
2- Institute of Physics, Meiji Gakuin University, Yokohama, Japan \\
3- High Energy Accelerator Research Organization (KEK), Tsukuba, Japan \\
4- Department of Physics, University of Toyama, Toyama, Japan 
}

\maketitle

\begin{abstract}
The Littlest Higgs Model with T-parity is one of the attractive candidates of physics beyond the Standard Model. One of the important predictions of the model is the existence of new heavy gauge bosons, where they acquire mass terms through the breaking of global symmetry necessarily imposed on the model. The determination of the masses are, hence, quite important to test the model. In this paper, the measurement accuracy of the heavy gauge bosons at the international linear collider (ILC) is reported.
\end{abstract}

\section{Introduction}
There are a number of scenarios for new physics beyond the Standard Model. The most famous one is the supersymmetric scenario. Recently, alternative one called the Little Higgs scenario has been proposed \cite{Arkani-Hamed:2001nc, Arkani-Hamed:2002qy}. In this scenario, the Higgs boson is regarded as a pseudo Nambu-Goldstone boson associated with a global symmetry at some higher scale. A $Z_2$ symmetry called T-parity is imposed on the models to satisfy constraints from electroweak precision measurements \cite{Cheng:2003ju, Cheng:2004yc, Low:2004xc}. Under the parity, new particles are assigned to be T-odd (i.e. with a T-parity of $-1$), while the SM particles are T-even. The lightest T-odd particle is stable and provides a good candidate for dark matter. In this article, we focus on the Littlest Higgs model with T-parity as a simple and typical example of models implementing both the Little Higgs mechanism and T-parity. 

In order to test the Little Higgs model, precise determinations of properties of Little Higgs partners are mandatory, because these particles are directly related to the cancellation of quadratically divergent corrections to the Higgs mass term. In particular, measurements of heavy gauge boson masses, Little Higgs partners for gauge bosons, are quite important. Since heavy gauge bosons acquire mass terms through the breaking of the global symmetry, precise measurements of their masses allow us to determine the most important parameter of the model, namely the vacuum expectation value of the breaking ($f$). 

We studied the measurement accuracy of masses of the heavy gauge bosons at ILC. In this article, the status of the study is reported.

\section{Representative point and target mode}
In order to perform a numerical simulation at ILC, we need to choose a representative point in the parameter space of the Littlest Higgs model with T-parity. Firstly, the model parameters should satisfy the current electroweak precision data. We also have satisfy the cosmological observation of dark matter relics. Thus, we consider not only the electroweak precision measurements but also the WMAP observation \cite{Komatsu:2008hk} to choose a point in the parameter space. We have selected a representative point where the Higgs mass and vacuum expectation value ($f$) are 134 GeV and 580 GeV, respectively. At the representative point, we have obtained $\Omega_{\rm DM} h^2$ of 1.05. The masses of the heavy gauge bosons are ($M_{A_{\mathrm{H}}}$, $M_{W_{\mathrm{H}}}$, $M_{Z_{\mathrm{H}}}$) = (81.9 GeV, 368 GeV, 369 GeV), where \ah, \zh, and \wh~are the Little Higgs partners of a photon, Z boson, and W boson, respectively. Here, \ah~plays the role of dark matter in this model \cite{Hubisz:2004ft, Asano:2006nr}. Since all the heavy gauge bosons are lighter than 500 GeV, it is possible to generate them at ILC.

\begin{table}[t]
\center{
\begin{tabular}{|c||c|c|c|c|}
\hline
$\sqrt{s}$ &
$e^+e^- \rightarrow A_{\mathrm{H}}Z_{\mathrm{H}}$ &
$e^+e^- \rightarrow Z_{\mathrm{H}}Z_{\mathrm{H}}$ &
$e^+e^- \rightarrow W_{\mathrm{H}}^+W_{\mathrm{H}}^-$ \\
\hline
 500 GeV & 1.31 (fb) & --- & --- \\
 \hline
 1 TeV & 6.99 (fb) & 99.5 (fb) & 233 (fb) \\
 \hline
 \end{tabular}
}
  \caption{Cross sections for the production of heavy gauge bosons at ILC.}
  \label{table:Xsections}
\end{table}

\begin{wrapfigure}{r}{0.35\columnwidth}
\centerline{\includegraphics[width=0.32\columnwidth]{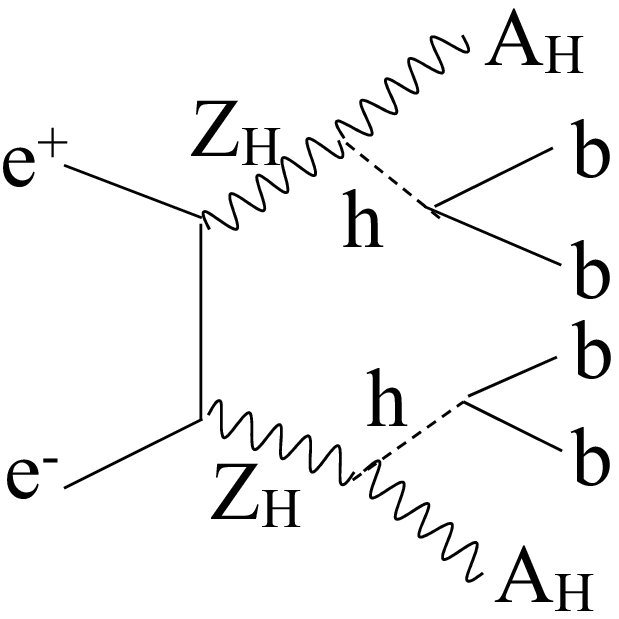}}
\caption{Diagrams for $e^{+} e^{-} \to Z_{\mathrm{H}} Z_{\mathrm{H}} \to A_{\mathrm{H}} A_{\mathrm{H}} bbbb$.}
\vspace{-0.8cm}
\label{fig:diagram}
\end{wrapfigure}

There are four processes whose final states consist of two heavy gauge bosons: $e^+e^- \rightarrow$ $A_{\mathrm{H}}A_{\mathrm{H}}$, $A_{\mathrm{H}}Z_{\mathrm{H}}$, $Z_{\mathrm{H}}Z_{\mathrm{H}}$, and $W_{\mathrm{H}}^+ W_{\mathrm{H}}^-$. The first process is undetectable because \ah~in the final state is a dark matter. The cross sections of the other processes are shown in Table \ref{table:Xsections}. We have studied \zhah~at $\sqrt{s} = 500$ GeV and \whwh~at $\sqrt{s} = 1$ TeV in the previous study \cite{prd}. In this article, we concentrate on evaluating the measurement accuracy of \zh~and \ah~by using \zhzh~at $\sqrt{s} = 1$ TeV, where \zh~decays into $A_{\mathrm{H}} h$ with almost 100\% branching fractions. We use $Z_{\mathrm{H}} Z_{\mathrm{H}} \to A_{\mathrm{H}} A_{\mathrm{H}} hh \to A_{\mathrm{H}}  A_{\mathrm{H}} bbbb$ as the signal mode. Feynman diagrams for the signal processes are shown in Fig. \ref{fig:diagram}. 

\section{Simulation tools}
We use Physsim \cite{physsim} to generate \zhzh~and all the standard model events, where the initial- and final-state radiation, beamstrahlung, and the beam energy spread are taken into account. The beam energy spread is set to 0.14\% for the electron beam and 0.07\% for the positron beam. The finite crossing angle between the electron and positron beams was assumed to be 14 mrad.. In Physsim, the helicity amplitudes were calculated using the HELAS library \cite{helas}, which allows us to deal with the effect of gauge boson polarizations properly. 
Parton showering and hadronization have been carried out by using PYTHIA6.4 \cite{pythia}, where final-state tau leptons are decayed by TAUOLA \cite{tauola} in order to handle their polarizations correctly. The generated Monte Carlo events have been passed to a detector simulator called JSFQuickSimulator, which implements the GLD geometry and other detector-performance related parameters \cite{glddod}.

\section{Analysis}
\begin{wrapfigure}{r}{0.6\columnwidth}
\centerline{\includegraphics[width=0.57\columnwidth]{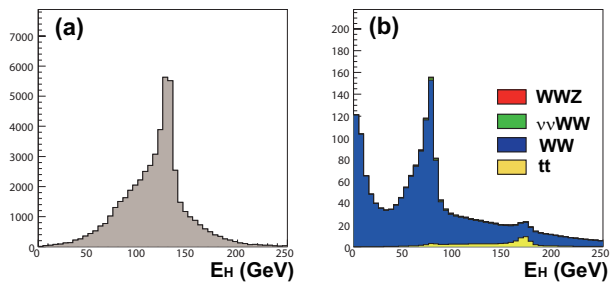}}
\caption{Distributions of the reconstructed Higgs mass for the signal (a) and background (b).}
\label{fig:hmass}
\end{wrapfigure}

Since \zhzhahahbbbb~is the signal mode in this analysis, all the events are reconstructed as 4-jet events. Then, two Higgs masses are reconstructed by using the $\chi^{2}$ function defined as 
\begin{equation}
\chi^{2} = \frac{(M_{\mathrm{H1}} - M_{\mathrm{H}})^{2}}{\sigma_{\mathrm{H}}^{2}} + \frac{(M_{\mathrm{H2}} - M_{\mathrm{H}})^{2}}{\sigma_{\mathrm{H}}^{2}},
\end{equation}
where $ M_{\mathrm{H1(2)}}$ is the reconstructed Higgs mass, $ M_{\mathrm{H}}$ is the Higgs mass (134 GeV), and $\sigma_{\mathrm{H}}$ is the Higgs mass resolution. Figure \ref{fig:hmass} shows distributions of the reconstructed Higgs mass for the signal and background. Since many background events contaminate in the signal region, we apply the selection cut to obtain the signal significance.

At first, we require the $\chi^{2}$ value below 80 to select the well reconstructed events. The signal events have 4 b-jets in the final states, whereas the dominant background, $WW$, has no b-jet. For that reason, we apply the b-tagging to reject $WW$ background. The requirement of the b-tagging is existence of 2 tracks with 3 $\sigma$ displacement from the interaction point, where $\sigma$ is the impact parameter resolution. We select events with the number of the b-tagged jets above 1. After the b-tagging, the dominant background becomes $t\bar{t}$. We investigated the acoplanarity ($\pi - \phi$), where $\phi$ is the angle between two reconstructed Higgs candidates in the plane perpendicular to the beam axis. Since the signal has large missing momentum due to \ah, the acoplanarity becomes large, comparing to $t\bar{t}$. We, therefore, select the acoplanarity above 25 degrees. The number of the events at each selection cut is summarized in Table \ref{tb:cut}.

\begin{table}[tb]
\center{
\begin{tabular}{|l|r|r|r|r|r|}
\hline
& \zhzh & $WWZ$ & $\nu \nu WW$ & $WW$ & $t\bar{t}$ \\ \hline
Xsec (fb) & 99.5 & 63.9 & 14.7 & 3,931 & 192.9 \\ \hline
No cut & 49,760 & 31,933 & 7,336 & 1,915,475 & 96,450 \\ \hline
$\chi^{2}<80$ & 41,789 & 8,763 & 1,785 & 226,783 & 65,911 \\ \hline
$N_{b} \geq 2$ & 30,442 & 1,824 & 151 & 6,866 & 64,906 \\ \hline
Acop. $>$ 25 deg. & 23,432 & 612 & 127 & 1,022 & 7,819 \\ \hline
\end{tabular}
}
\caption{Cut summary.}
\label{tb:cut}
\end{table}

The masses of \ah~and \zh~bosons can be extracted from the edges of the distribution 
of the reconstructed Higgs boson energies. 
This is because the maximum and minimum Higgs boson energies 
($E_{\mathrm{max}}$ and $E_{\mathrm{min}}$) are written in terms of these masses,

\begin{eqnarray}
 E_{\mathrm{max}}
 &=& 
 \gamma_{Z_{\mathrm{H}}} E^{\ast}_{h}
 + 
 \beta_{Z_{\mathrm{H}}} \gamma_{Z_{\mathrm{H}}} p^{\ast}_{h},
 \nonumber \\ 
 E_{\mathrm{min}}
 &=& 
 \gamma_{Z_{\mathrm{H}}} E^{\ast}_{h}
 - 
 \beta_{Z_{\mathrm{H}}} \gamma_{Z_{\mathrm{H}}} p^{\ast}_{h},  
 \label{eq:eedge}
\end{eqnarray}
where $\beta_{Z_{\mathrm{H}}} (\gamma_{Z_{\mathrm{H}}})$ is the $\beta (\gamma)$ factor 
of the \zh~boson in the laboratory frame, while $E^{\ast}_{h} 
(p_{h}^{\ast})$ is the energy (momentum) of the Higgs boson 
in the rest frame of the \zh~boson. Note that $E^{\ast}_{h}$ is given 
as $(M_{Z_{\mathrm{H}}}^2 + M_h^2 - M_{A_{\mathrm{H}}}^2)/(2M_{Z_{\mathrm{H}}})$.

\begin{wrapfigure}{r}{0.6\columnwidth}
\centerline{\includegraphics[width=0.57\columnwidth]{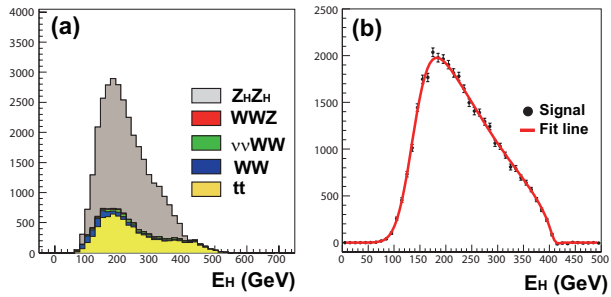}}
\caption{(a)Energy distribution of the reconstructed Higgs bosons 
with remaining backgrounds after the selection cuts. (b) Energy distribution of the Higgs bosons after subtracting the backgrounds. The distribution is fitted by a line shape function determined with a high statistics signal sample.}
\label{fig:massfit}
\end{wrapfigure}

Figure \ref{fig:massfit}(a) shows the energy distribution of the reconstructed Higgs bosons with remaining backgrounds. The background events are subtracted from Fig. \ref{fig:massfit}(a), assuming that the background distribution can be understand completely. Then, the endpoints, $E_{\mathrm{max}}$ and $E_{\mathrm{min}}$, have been estimated by fitting the distribution with a line shape determined by a high statistics signal sample. The fit resulted in $m_{A_{\mathrm{H}}}$ and $m_{Z_{\mathrm{H}}}$ to be $82.0 \pm 3.5$ GeV and $367.2 \pm 3.5$ GeV, respectively, which should be compared to their true values: 81.85 GeV and 368.2 GeV. From this result, the masses of \ah~and \zh~can be determined with accuracies of 4.8\% and 0.9\%, respectively.

\section{Summary} \label{sec:summary}
The Littlest Higgs Model with T-parity is one of the attractive candidates of physics beyond the Standard Model since it solves both the little hierarchy and dark matter problems simultaneously. One of the important predictions of the model is the existence of new heavy gauge bosons, where they acquire mass terms through the breaking of global symmetry necessarily imposed on the model. The determination of the masses are, hence, quite important to test the model. 

We have performed Monte Carlo simulations in order to estimate measurement accuracy of the masses of the heavy gauge bosons, \ah~and \zh, at ILC with $\sqrt{s} = 1$ TeV. After the selection cuts, we extract the masses of \ah~and \zh, fitting the energy distribution of Higgs bosons. The masses of \ah~and \zh~can be determined with accuracies of 4.8\% and 0.9\%, respectively. In the previous study, the measurement accuracy of \ah~mass was 1.4\% by using \whwh~at $\sqrt{s} = 1$ TeV. By the simultaneous fitting with \whwh~and \zhzh, we will obtain better mass resolution for \ah.

\section{Acknowledgments}
The authors would like to thank all the members of the ILC physics subgroup \cite{softg} for useful discussions. This study is supported in part by the Creative Scientific Research Grant No. 18GS0202 of the Japan Society for Promotion of Science, and Dean's Grant for Exploratory Research in Graduate School of Science of Tohoku University.


\begin{footnotesize}


\end{footnotesize}



\begin{thebibliography}{99}
\bibitem{Arkani-Hamed:2001nc}
  N.~Arkani-Hamed, A.~G.~Cohen and H.~Georgi,
  Phys.\ Lett.\ B {\bf 513} (2001) 232;

\bibitem{Arkani-Hamed:2002qy}
  N.~Arkani-Hamed, A.~G.~Cohen, E.~Katz and A.~E.~Nelson,
  JHEP {\bf 0207} (2002) 034.

\bibitem{Cheng:2003ju}
  H.~C.~Cheng and I.~Low,
  JHEP {\bf 0309} (2003) 051.

\bibitem{Cheng:2004yc}
  H.~C.~Cheng and I.~Low,
  JHEP {\bf 0408} (2004) 061.

\bibitem{Low:2004xc}
  I.~Low,
  JHEP {\bf 0410} (2004) 067.

\bibitem{Komatsu:2008hk}
  E.~Komatsu {\it et al.}  [WMAP Collaboration],
  arXiv:0803.0547 [astro-ph].

\bibitem{Hubisz:2004ft}
  J.~Hubisz and P.~Meade,
  Phys.\ Rev.\ D {\bf 71} (2005) 035016,
  (For the correct paramter region consistent with the WMAP observation,
   see the figure in the revised vergion, hep-ph/0411264v3).

\bibitem{Asano:2006nr}
  M.~Asano, S.~Matsumoto, N.~Okada and Y.~Okada,
  Phys.\ Rev.\  D {\bf 75} (2007) 063506;

\bibitem{prd} E. Asakawa {\it et al.}, Phys. Rev. D79, 075013, (2009).
\bibitem{physsim} http://acfahep.kek.jp/subg/sim/softs.html.
\bibitem{helas} H. Murayama, I. Watanabe, K. Hagiwara, KEK-91-11, (1992) 184.
\bibitem{pythia} T. Sj$\dot{\mathrm{o}}$strand, \emph{Comp, Phys. Comm.} {\bf 82} (1994) 74.
\bibitem{tauola} http://wasm.home.cern.ch/wasm/goodies.html.
\bibitem{glddod} GLD Detector Outline Document, arXiv:physics/0607154.
\bibitem{softg} http://www-jlc.kek.jp/subg/physics/ilcphys/.

\end{thebibliography}
\end{document}